\documentclass[twocolumn,prl]{revtex4} 
\usepackage{epsfig}
\usepackage{color} 
\providecommand{\tabularnewline}{\\}
\begin{document} 
\title{Quantum Friction in Nanomechanical Oscillators at Millikelvin Temperatures}  
\author{Guiti Zolfagharkhani$^1$, Alexei Gaidarzhy$^{1,2}$, Seung-Bo Shim$^{1,*}$, Robert L. Badzey$^1$, and Pritiraj Mohanty$^1$} 
\affiliation{$^1$Department of Physics, Boston University, 590 Commonwealth Avenue, Boston, MA 02215}    
\affiliation{$^2$Aerospace and Mechanical Engineering, Boston University, 110 Cummington Street, Boston, MA 02215}

\makeatletter
\global\@specialpagefalse
\def\@oddhead{\hskip 5.0in { Phys. Rev. B {\bf 72}, 224101 (2005)}}
\let\@evenhead\@oddhead
\makeatother

%\maketitle
\begin{abstract} 
We report low-temperature measurements of dissipation in megahertz-range, suspended, single-crystal nanomechanical oscillators. At millikelvin temperatures, both dissipation (inverse quality factor) and shift in the resonance frequency display reproducible features, similar to those observed in sound attenuation experiments in disordered glasses and consistent with measurements in larger micromechanical oscillators fabricated from single-crystal silicon. Dissipation in our single-crystal nanomechanical structures is dominated by internal quantum friction due to an estimated number of roughly 50 two-level systems, which represent both dangling bonds on the surface and bulk defects.  

\vskip 0.2 in  
\end{abstract}  
\maketitle

Understanding dissipation in nanomechanical oscillators is essential to their study as well as to their applications that are of fundamental and technical interests \cite{blick,bishop,weiss}.  Applications of classical micro- and nano-mechanical systems---for instance, as single spin detectors \cite{rugar} and spin-torque oscillators \cite{fulde}, memory elements \cite{badzey}, or as ultrasensitive sensors of biological interactions \cite{wu} or fundamental forces such as gravity \cite{long,chiaverini}---will be limited by energy dissipation, usually quantified in terms of the quality factor (Q). The realization of a mechanical quantum system and its coherent manipulation will depend on the limitations imposed by decoherence and dissipation. There is a further motivation in exploring mechanical and elastic effects in nanomechanical structures of mesoscopic system size; specifically, in multi-scale modelling of mechanical systems consisting of around 100 million atoms, which can now be fabricated and experimentally characterized comprehensively \cite{kalia}.  

Central to the study of dissipation is the identification of the dominating dissipation mechanism and its scaling with system size. Some of the identifiable mechanisms are energy loss due to clamping, thermoelastic processes, gas friction, mode anharmonicity, metal film friction, and coupling to dislocations and localized internal defects. Since nanomechanical systems possess large surface-to-volume ratios, these mechanisms are treated essentially as surface effects. Of particular interest is the study of energy dissipation or internal friction by quantum tunneling of two-level systems that become effective typically below a temperature of 1 Kelvin. Despite theoretical interest, experimental studies of internal friction in nanomechanical systems at millikelvin temperatures are yet to be reported.
 
Kleiman et al. \cite{kleiman} have measured dissipation in macroscopic centimeter-scale single-crystal silicon resonators at millikelvin temperatures. The temperature dependence is similar to that in vitreos silica, which can be essentially understood in the framework of the so-called glass model of two-level systems \cite{phillips1,parpia1,pohl}. Their results were  subsequently reinterpreted by Phillips in light of asymmetry distribution produced by local strains in crystals \cite{phillips2}. Greywall et al. \cite{greywall} have identified an anomalous contribution to the low temperature dissipation in single-crystal silicon resonators in the form of distinct periodic peaks. Carr et al., \cite{parpia2} Olkhovets et al., \cite{craighead} Evoy et al., \cite{evoy} and Carr and Craighead \cite{carr} have characterized intrinsic dissipation in nano-electro-mechanical resonators (NEMS) in the 1-10 MHz range, identified with ``surface and near-surface phenomena''.  Yang et al. \cite{yang} have studied the surface contributions to dissipation in ultrathin single-crystal silicon cantilevers. However, none of the nano-scale dissipation experiments were done at millikelvin temperatures.

In this Letter, we report detailed measurements of frequency shift($\delta f/f$) and dissipation ($Q^{-1}$) of a set of suspended megahertz-range nanomechanical oscillators at temperatures down to 60 mK. We extract frequency shift and dissipation from the Lorentzian response of the resonators by driving them on resonance in the linear regime. The real and imaginary parts of the susceptibility function $\chi(\omega_0)=\chi^\prime(\omega_0)+i\chi^{\prime\prime}(\omega_0)\simeq \rho v^2 (2 \delta f/f_0 + iQ^{-1})$ are proportional to $\delta f/f$ and $Q^{-1}$ at frequencies close to the resonance frequency \cite{esquinazi}. The susceptibility function quantifies the system response due to its coupling to an environment, which results in a loss of both energy (dissipation) and quantum coherence (decoherence). Further motivation for doing measurements on resonance is to enhance the sensitivity of system response to small dissipative forces.

Both the relative shift in resonance frequency $\delta f/f$ and dissipation $Q^{-1}$, quantified by the full width at half maximum (FWHM) of the resonance peak, demonstrate reproducible temperature dependence. We find that the temperature dependence of frequency shift follows the expected logarithmic behaviour down to the lowest measured temperature of 60 mK without saturation. This is consistent with the Phillips model of disordered crystals and the glass model of tunneling two-state systems. On the other hand, the temperature dependence of dissipation does not show the expected T or $T^3$ dependence \cite{esquinazi}. Our single-crystal structures act essentially as surface systems due to their large surface-to-volume ratio. This results in a broad distribution of tunnel-splitting energies or glass-like behavior. Furthermore, by quantitative comparison to the glass model, we find that the dissipation in our nanomechanical structures is due to the internal friction of roughly 50 two-state atoms and dangling bonds. However, at temperatures below 100 mK, the structures demonstrate excess temperature-independent dissipation, which cannot be explained with either the glass or the Phillips model \cite{phillips2}.
  
We fabricate our nanomechanical beams from  single crystal Si/SiO2/Si wafers using e-beam lithography and nanomachining. We have measured two different beams with lengths 6 and 7 $\mu$m (width 300 nm and thickness 200 nm for both beams). The top surface of the beams is covered with a thermally evaporated 10 nm Cr underlayer and 80 nm Au film. We measure the structures magnetomotively at the center of a 16 Tesla superconducting magnet in a dilution refrigerator. 
%____________________________________________________________________ 
\begin{figure}[t] 
\epsfxsize=8.7 cm 
\epsfysize=6 cm 
\epsfbox{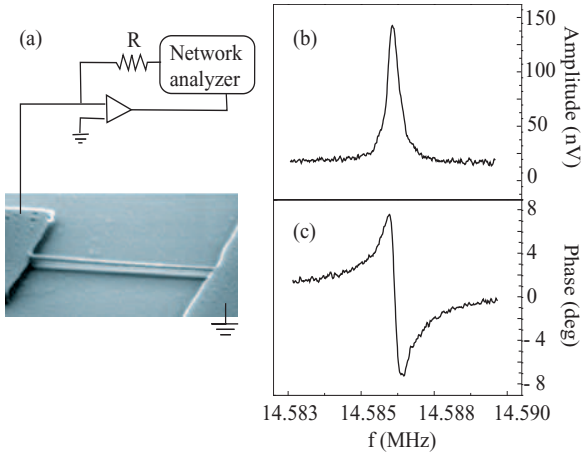} 
\caption{
(Color online) a) SEM micrograph of the 7 $\mu$m nanobeam. The diagram shows the measurement circuit.  To drive the beam, we pass an alternating current $I(\omega)$ across the beam of length $L$ in the presence of magnetic field $B$, to set up a force $F_{dr}(\omega)= I(\omega)LB$ that is orthogonal to both. The center beam displacement $x$ for the fundamental mode shape in the magnetic field $B$ induces a voltage $V_{emf}(\omega)=\xi LB\omega_0x(\omega)$ in the gold electrode. We detect the response voltage using a RF network analyzer. b) The fundamental resonance mode of the beam at 14.586 MHz with $Q = 33000$. The induced voltage $V_{emf}$ is a Lorentzian peak on top of a white noise background, which is due primarily to the input noise of our preamplifier, and c) the corresponding phase.  
}  
\end{figure}  
%_________________________________ 
%_____________________________________________________________________ 
\begin{figure}[t] 
\epsfxsize=8.7 cm 
\epsfysize=7 cm 
\epsfbox{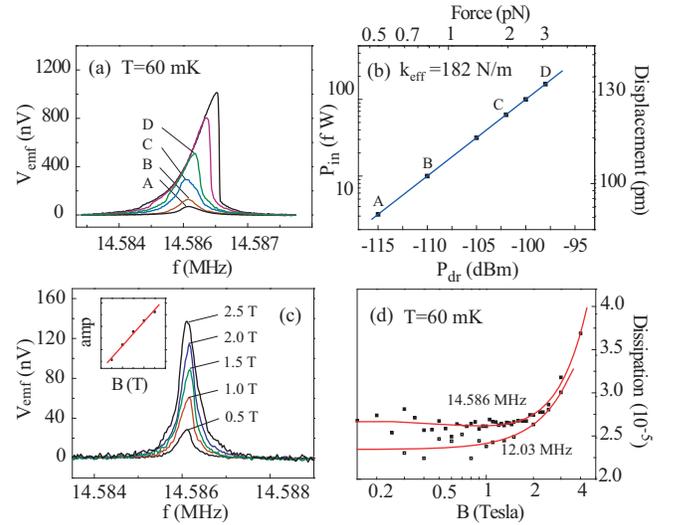} 
\caption{
(Color online) a) The resonance peak, which is Lorentzian in the linear regime, assumes the asymmetric nonlinear shape when the driving power is increased  above -95 dBm (trace D on the plot). b) Linear dependence of the induced power $P_{in}$ on driving power $P_{dr}$ at 4 Tesla. The induced power corresponds to the resonance displacement $x$, shown versus the driving force on a log-log scale. The measured effective spring constant is $k_{eff} = 182\mbox{ N/m}$. c) The response of the oscillator increases linearly with magnetic field, as expected when the driving force $F=ILB$ is kept constant at 20 pN. We keep the driving force constant in order to investigate the effect of the magnetic field on energy dissipation. d) Magnetic field dependence of dissipation. The data set  is taken at 60 mK. It exhibits quadratic field dependence for high magnetic fields in both samples. Dissipation saturates in the low field region. 
}  
\end{figure}  
%_____________________________________________________________________ 

We observe the fundamental resonance modes of the two beams at frequencies 12.028 and 14.586 MHz. A typical linear response of the beam at 14.586 MHz with quality factor Q=33000 is shown in Fig.~1b. Dissipation is defined as the inverse of the quality factor $Q=\omega / \Delta\omega$. The displacement of the beam on resonance varies linearly with the driving force according to Hooke's Law $x(\omega_0)=Q F_{dr}(\omega_0)/k_{eff}$\cite {mohanty}. The effective spring constant $k_{eff} = 182\mbox{ N/m}$ is inversely proportional to the linear fit in Fig.~2b.

The response $V_{emf}$ on resonance $\omega \simeq \omega_0$ is given by 
$V_{emf}(\omega_0) = {\xi L^2 B^2 Q\over {m\omega_0}}I_{dr} (\omega_0)$, 
according to the magnetomotive scheme. We measure the induced voltage of the response at constant force $F_{dr}= ILB= 20 \mbox{ pN}$ (Fig.~2c) to verify the linear $B$ dependence of the response. Fig.~2d shows magnetic field dependence of dissipation in two samples at 60 mK. The quadratic dependence of dissipation suggests the presence of charged defects or impurities in the samples. The scattering rate of the charged impurities is adjusted by the magnetic field and can be shown to give quadratic dependence \cite{mohanty}. Below 2 Tesla, dissipation saturates with respect to the field in both samples. The following temperature measurements are done at the saturation field of 2 Tesla. The field dependence of dissipation was measured at different temperatures up to 144 mK and showed the same quadratic behavior.
%_____________________________________________________________________ 

\begin{table}
\caption{Quality factor from the different damping mechanisms. Here $L=6$ $\mu$m, $w=300$ $nm$ and $t=200$ $nm$ are the length, width and tickness of the beam and $h=675$ $\mu$m is the thickness of the substrate.}
\makeatother
\begin{tabular}{|l|l|}

\multicolumn{2}{c}{Quality factor}\tabularnewline
\hline
Clamping\cite{photiadis}&
$Lh^{2}/0.95wt^{2}=2.5\times10^{8}$\tabularnewline
\hline
Clamping\cite{photiadis} $h\gg t$ &
$L^{5}/0.31wh^{4}=5\times10^{7}$\tabularnewline
\hline
Constriction\cite{cross} &
$L/w=20$\tabularnewline
\hline
Thermoelastic\cite{lifshitz} at 10 K&
 $10^{10}$\tabularnewline
\hline
\end{tabular}
\end{table}
%_____________________________________________________________________ 

Table 1 summarizes calculation results found in literature for various clamping conditions and thermoelastic damping. Clamping loss is due to the strain energy radiating into the support structure of the suspended resonator. Cross and Lifshitz \cite{cross} have estimated the elastic wave transmission across an abrupt junction in a one dimensional scalar model. Their assumption of irreversible energy loss into the supporting pads results in a overestimate of dissipation $Q^{-1}=w/L$. Photiadis and Judge \cite{photiadis} extended the analysis of the doubly-clamped beam and calculated the clamping loss. Using our beam parameters, the resulting energy loss due to clamping is found to be on the order of $10^{-8}$. Thermolastic damping is due to the interaction of the normal modes of vibration of a resonator with thermally excited elastic modes. Lifshitz and Roukes \cite{lifshitz} have estimated thermoelastic damping for silicon in a regime where thermal phonons are diffusive. The value of $Q^{-1}$ they report is less than $10^{-10}$ at temperatures below 10 Kelvin.  The observed dissipation in our beams is orders of magnitude greater than both clamping and thermoelastic loss.  Additional friction and self-heating arises from the metal electrode on the top surface of our beams, as shown in the experiments of Ref. 27. The latter mechanism is a nonlinear effect that becomes inapplicable in our linear response regime, where the dissipation is independent of the driving current.  We verify this fact by monitoring the response peak width at different driving voltages in the linear regime (not shown here) and observing no variation, within the measurement error.  Furthermore, the temperature dependence of the data below 1 Kelvin is not expected to result from either of these mechanisms. We rule out thermoelastic and clamping loss as well as metal electrode friction as mechanisms responsible for the observed dissipation in our oscillators. 
%_____________________________________________________________________
\begin{figure}[t] 
\epsfxsize=8.7 cm 
\epsfysize=3.6 cm 
\epsfbox{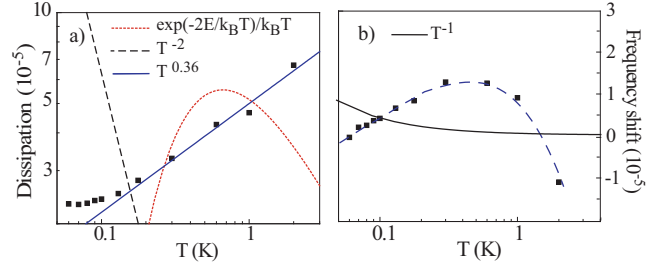} 
\caption{
(Color online) a) Temperature dependence of dissipation for the 12 MHz beam on a log-log scale. The $T^{-2}$ and $exp(-2e/k_BT)/k_BT$ curves are the predicted acoustic dissipation forms due to TLS in crystals at low and high temperatures respectively. The best fit to our dissipation data is the $T^{0.36}$ power law.  b) Temperature dependence of the shift in the resonance frequency on a semi-log scale with a guide to the eye (dashed line). According to the acoustic dissipation in crystals, the temperature dependence of the frequency shift at higher temperatures goes as $T^{-1}$. 
}  
\end{figure}  
%_____________________________________________________________________ 

%______________________________________________________________
\begin{table}
\caption{Temperature dependence of the dissipation and shift in the resonance frequency arising from the glass and single crystal models. Here $C=n_0D^2/\rho v^2$, E is the energy of TLS, $a=\pi^3 C M^2 k^3/24 v^5 \rho \hbar^4 \omega$, $b=2C\omega D/(\omega_0^2+D^2)$ and $d=\omega /k_B^2D$. The number of TLS and the density of states for the TLS ($J^{-1}m^{-3}$) for both models are given.} 
\makeatother
\begin{tabular}{|l|l|l|l|l|l|}

\multicolumn{5}{c}{Two Level Systems}\tabularnewline
\hline 

\multicolumn{3}{|c|}{Glass Model}&
\multicolumn{2}{c|}{Single Crystal Model}\tabularnewline
\cline{2-3} \cline{4-5} 
\hline 
&
\small{$T<T_{*}$}&
\small{$T>T_{*}$}&
\small{$E>k_B T$}&
\small{$E<k_B T$}\tabularnewline
\hline 
$\delta f/f$&
\small{$Cln(T/T_{0})$}&
\small{$-{1\over 2}Cln(T/T_{0})$}&
\small{$C/E$}&
\small{$C/2k_BT$}\tabularnewline
\hline 
$Q^{-1}$&
$aT^{3}$&
$1/2\pi C$&
\small{$be^{-2E/k_BT}/k_BT$}&
$dT^{-2}$\tabularnewline
\hline 
n&
\multicolumn{2}{c|}{50}&
\multicolumn{2}{c|}{$0.13$}\tabularnewline
\hline
$n_0$ &
\multicolumn{2}{c|}{$10^{44}$ \small{$(J^{-1} m^{-3})$}}&
\multicolumn{2}{c|}{$3\times 10^{41}$ \small{$(J^{-1} m^{-3})$}}\tabularnewline
\hline

\multicolumn{3}{|c|}{Phillips Model}
&\multicolumn{2}{c|}{}\tabularnewline
\hline
&
\small{$E>k_B T$}&
\small{$E<k_B T$}
&\multicolumn{2}{c|}{}\tabularnewline
\hline 
$\delta f/f$&
\small{$\propto ln(T/T_{0})$}&
\small{$\propto -ln(T/T_{0})$}
&\multicolumn{2}{c|}{}\tabularnewline
\hline 
$Q^{-1}$&
\small{$\propto T$}&
\small{$Constant$}
&\multicolumn{2}{c|}{}\tabularnewline
\hline 
n&
\multicolumn{2}{c|}{---}
&\multicolumn{2}{c|}{}\tabularnewline
\hline
$n_0$ &
\multicolumn{2}{c|}{---}
&\multicolumn{2}{c|}{}\tabularnewline
\hline

\end{tabular}
\end{table}
%______________________________________________________________

Dissipation behavior very similar to what we observe (Fig.~4) has been seen for the acoustical properties and the dissipation in amorphous materials and disordered crystals \cite{phillips1,phillips2,wurger,classen}. It can be described by the existence of two-level systems (TLS), which represent defects in the crystal. The defects arise from the intrinsic impurity atoms in the bond structure, the broken or dangling bonds on the surface due to the abrupt termination of the crystal structure, and contamination by other atoms like oxygen and water molecules. According to the glass model, these TLS move in asymmetric double-well potentials with asymmetry $\Delta$ and characteristic tunnel splitting energies $\Delta_0$ on the order of the thermal energy, or $10^{-4} \mbox{ eV}$ at 1K.  The TLS couple to the phonons of the lattice resonantly at low temperatures (typically below 1K), while at higher temperatures the phonons modulate the asymmetry energies which results in relaxation absorption by the TLS \cite{kleiman,esquinazi,mohanty}. In glasses the distribution of the asymmetry energies is very broad and the density of states $P$ for the TLS's is taken to be constant in terms of the parameters $\Delta$ and $\Delta_0$, $P(\Delta,\Delta_0)d \Delta d \Delta_0 = {P_0 \over \Delta_0} d \Delta d \Delta_0 $ \cite{esquinazi}, whereas for single crystals Phillips has suggested that similar thermal properties can be obtained with a well defined tunnel splitting energy and a Gaussian asymmetry distribution of width $\Delta_1$, $P(\Delta) d\Delta = A exp(- {\Delta^2 / 2\Delta_1^2}) d\Delta$ \cite{phillips2}.  Both approaches agree on the low temperature behavior of frequency shift and dissipation, and the dependences are outlined in Table 2.  

In single crystals, dissipation due to TLS is expected to have a $T^{-2}$ form at high temperatures  $E<k_B T$, and decreases as $T^{-1}e^{-2E/k_B T}$ at low temperatures $E>k_B T$ where $E$ is the energy of TLS. The sound velocity is predicted to follow the $T^{-1}$ dependence at high temperatures, saturating at lower temperatures \cite{mohanty,wurger,classen}. Fig.~3 shows poor agreement of the crystalline impurity dissipation theory with our data. 

%______________________________________________________________________  
 \begin{figure}[t] 
\epsfxsize=8.7 cm 
\epsfysize=8 cm 
\epsfbox{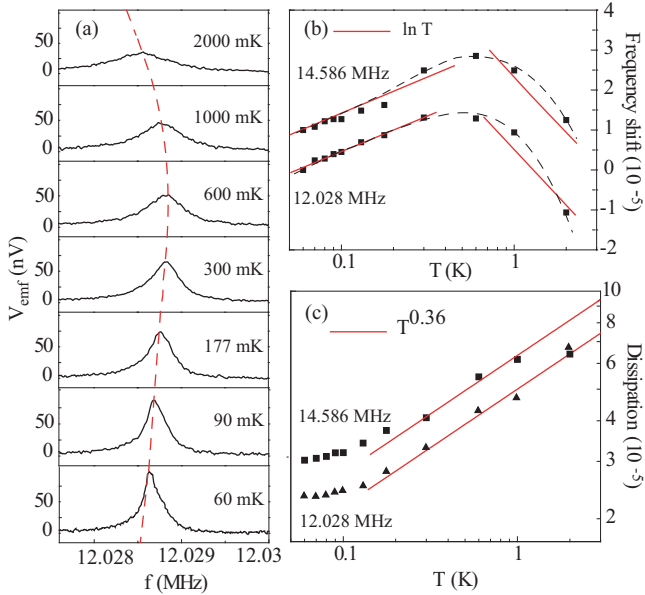} 
\caption{
(Color online) a) The sharp response of the 12 MHz beam as temperature is increased displays significant frequency shift and dissipation. The shift in the resonance frequency is illustrated by means of a guide to the eye (dashed line), drawn on the raw data. b) Temperature dependence of the shift in the resonance frequency for both beams on a semi-log scale. The plot for the 14.586 MHz beam has been shifted up for clarity. The data exhibits a peak in the vicinity of 0.7 K in both samples. Below the temperature 0.7 K the dependence is closely logarithmic lnT. However, it is difficult to determine the behavior above 0.7 K due to the sparseness of the data. c) Below 2 Kelvin dissipation shows weak power law dependence $T^{0.36}$. Both samples exhibit saturation of dissipation below 100 mK.}  
\end{figure}  
%_____________________________________________________________________ 

In Fig.~4 we observe logarithmic temperature dependence in the frequency shift data corresponding to resonant absorption at temperatures below $T_0=0.7 \mbox{ K}$. The temperature dependence of dissipation in our data $T^{0.36}$ is weaker than the expected $T$ or $T^{3}$ dependence due to TLS relaxation via electron or phonon channels respectively \cite{phillips2,hunklinger}, or the T dependence in the Phillips model for a crystal with impurities. The contribution of resonant interactions to the frequency shift is given by
\begin{equation}
{\delta f \over f_0} = C \ln{T \over T_0}; \quad \quad T<T_0,        
\end{equation} 
where $T_0=60 \mbox{ mK}$ is a reference temperature \cite{thompson}. C is defined by $C={n_0D^2 \over \rho v^2}$ where $\rho$ and $v$ are density and sound velocity, $n_0$ is the density of states for the TLS and D is the deformaion potential. We  have calculated the number of two-level systems in our silicon beams using $D=0.4 \mbox{ eV}$ \cite{hunklinger}, $\rho =2330\mbox{ kg/m}^3$ and $v=5000 \mbox{ m/s}$. We find that the measured dissipation is due to the coupling of the resonant phonon mode to as few as 50 atoms in the silicon beams.

Because of its small size in the micron/nano scale and the associated large surface-to-volume ratio, the single-crystal structure appears to behave like a disordered or amorphous system with a wide distribution of asymmetry energies due to surface roughness.  Although both the glass model as well as the Phillips model for disorderd crystals account for most of the features, the low temperature saturation of dissipation (with no corresponding saturation of the frequency shift) cannot be explained in this framework. This excess dissipation at low temperatures may be understood in a more sophisticated analysis of the TLS such as the Caldeira-Leggett model in the sub-ohmic regime \cite{leggett,paco}. 

In conclusion, we have measured the dissipation and frequency shift in the resonance frequency of MHz-range nanomechanical beams at millikelvin temperatures. The temperature dependence exhibits partial agreement with both glass and disorderd crystal models of tunneling two-level systems (TLS), with quantum dissipation arising from coupling to as few as 50 TLS.  A proper understanding of the quantum dissipation mechanisms in nanomechanical structures at millikelvin temperatures will be essencial to the realization of the quantum harmonic oscillator \cite{gaidarzhy,gaidarzhy2}. Our results help explain the apparent upper limit on the quality factor of nano-electro-mechanical structures (NEMS) with decreasing system size. Their smaller sizes make NEMS progressively susceptible to (quantum) dissipation by a handful of two-level systems arising from dangling or uncompensated bonds, which, even in the cleanest structures, are as unavoidable as the required termination of the surface.
This work is supported by the National Science Foundation (grant number DMR-0346707) under the NSF-EC Cooperative Activity in Materials Research.

*Current Address: CSCMR and School of Physics, Seoul National University, Seoul 151-747, South Korea
%_____________________________________________________________________ 

\end{document}